\documentclass[conference]{IEEEtran}
\IEEEoverridecommandlockouts

\usepackage{cite}
\usepackage{amsmath,amssymb,amsfonts}
\usepackage{algorithmic}
\usepackage{graphicx}
\usepackage{textcomp}
\usepackage{xcolor}

\newcommand{\tramlib}{{\em TramLib}}
\newcommand\charm{Charm++}
\newcommand\libtram{TramLib}
\newcommand\pingack{PingAck}
\newcommand\nn{{\em PP}}
\newcommand\pp{{\em WW}}

\newcommand\pns{{\em WPs}}
\newcommand\psn{{\em WsP}}

\begin{document}

\title{Shared Memory-Aware Latency-Sensitive Message Aggregation for Fine-Grained Communication}

\author{\IEEEauthorblockN{1\textsuperscript{st} Kavitha Chandrasekar}
\IEEEauthorblockA{\textit{Dept. of Computer Science} \\
\textit{Univ of Illinois at Urbana-Champaign}\\
Urbana, USA \\
kchndrs2@illinois.edu}
\and
\IEEEauthorblockN{2\textsuperscript{nd} Laxmikant Kale}
\IEEEauthorblockA{\textit{Dept. of Computer Science} \\
\textit{Univ of Illinois at Urbana-Champaign}\\
Urbana, USA \\
kale@illinois.edu}
}

\maketitle

\makeatletter
\def\ps@IEEEtitlepagestyle{
  \def\@oddfoot{\copyrightnotice}
  \def\@evenfoot{}
}
\def\copyrightnotice{
  {\footnotesize
  \begin{minipage}{\textwidth}
  \centering
  Copyright~\copyright~\the\year{} IEEE.  Personal use of this material is permitted.  Permission from IEEE must be obtained for all other uses, in any current or future media, including reprinting/republishing this material for advertising or promotional purposes, creating new collective works, for resale or redistribution to servers or lists, or reuse of any copyrighted component of this work in other works.
  \end{minipage}
  }
}

\begin{abstract}
Message aggregation is often used with a goal to reduce communication cost in HPC applications. The difference in the order of overhead of sending a message and cost of per byte transferred motivates the need for message aggregation, for several irregular fine-grained messaging applications like graph algorithms and parallel discrete event simulation (PDES). While message aggregation is frequently utilized in ``MPI-everywhere" model, to coalesce messages between processes mapped to cores, such aggregation across threads in a process, say in MPI+X models or Charm++ SMP (Shared Memory Parallelism) mode, is often avoided. Within-process coalescing is likely to require synchronization across threads and lead to performance issues from contention. However, as a result, SMP-unaware aggregation mechanisms may not fully utilize aggregation opportunities available to applications in SMP mode.
Additionally, while the benefit of message aggregation is often analyzed in terms of reducing the overhead, specifically the per message cost, we also analyze different schemes that can aid in reducing the message latency, ie. the time from when a message is sent to the time when it is received. Message latency can affect several applications like PDES with speculative execution where reducing message latency could result in fewer rollbacks. To address these challenges, in our work, we demonstrate the effectiveness of shared memory-aware message aggregation schemes for a range of proxy applications with respect to messaging overhead and latency.
\end{abstract}

\section{Introduction}
\label{sec:intro}
Message aggregation libraries aim to provide an interface for applications for consolidating messages to reduce communication cost associated with processing many messages. Previous schemes for aggregation like TRAM~\cite{tramICPP14} in ~\charm{} and streaming library Active Pebbles~\cite{active-pebbles} use topology-aware routing schemes but these are less beneficial for modern topologies like fat-trees. 

Modern machines have many tens of cores per physical node (also called ``host"). Applications tend to use them in two modes: in one mode there is a separate process bound to each core that is used by the application; this is called MPI-everywhere for MPI applications, and non-SMP mode in Charm++ applications. The other mode uses multiple processes per node, but each process owns multiple cores, and facilitates shared memory parallelism (SMP) across cores within a process. For MPI applications, parallelism within a process and across cores is managed by one of the thread-parallel libraries such as OpenMP or Kokkos. Charm++ does not require a separate programming model for within-process communication, but utilizes the shared memory for optimizing inter-object communication and cooperation. It is this SMP mode that is the focus of our work. 
This SMP scenario amplifies the need for message aggregation further.  

For the MPI-everywhere mode, systems like YGM~\cite{ygm-smp} have recently looked into implementing node-aware schemes for message aggregation. YGM's mechanism for coalescing messages at node level uses a phase of node-local exchange which relies on underlying MPI implementation's method of achieving cheaper within node communication (eg. xpmem or cma). But the context we aim at is where worker threads divide the application's work among cores, but need to exchange many short messages with other workers, in a shared memory setting.

The SMP mode, in Charm++ as well as MPI, presents several potential advantages to the applications. Worker threads can share and divide work more efficiently, for better load balance (using dynamic schedules in OPENMP or stealable tasks and object migration in Charm++. Large read-only data structures can be shared among workers without making multiple copies. Even dynamic data structures, such as software particle (or tree) caches in astronomy codes can save duplicate external (across-node) communications. These orthogonal benefits may make SMP mode attractive independent of message-aggregation considerations. 

The questions we explore in this paper are: Given that an application wants to operate in SMP mode, what options are there for organzining message aggregation and how do they compare with each other? Does the SMP mode aggregation by itself provides benefits over non-SMP aggregation that makes SMP mode attractive even for applications that don't particularly require SMP mode? Finally, what metrics are important for such comparisons? 

\begin{figure}[h!]
  \centering
  \includegraphics{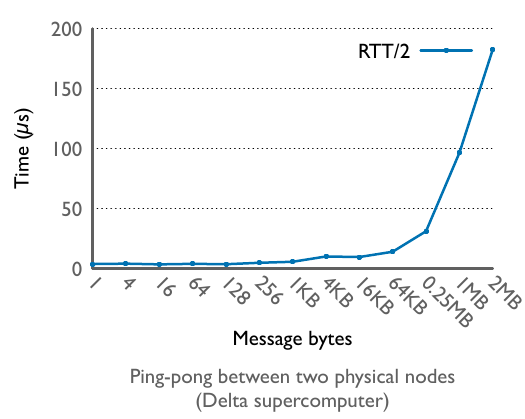}
  \caption{Time to send a message does not change for small byte count, suggesting time is dominated by latency $\alpha$ of $\mu$seconds}
  \label{fig:latency}
\end{figure}

The main metric that has motivated research on message aggregation has been the overhead. The cost associated with sending a message over network can be computed using the alpha-beta communication model as $\alpha + N \beta $, where N is the number of bytes. Here $\alpha$ is the latency per message and $\beta$ (inverse of bandwidth) is the cost per byte transferred. 
To illustrate the costs, we measure the time taken to send a message (roundtrip time/2) between two physical nodes on the Delta supercomputer using a ping-pong benchmark. As we can see in Figure~\ref{fig:latency}, the time for small message sizes is dominated by latency cost ($\alpha$) and is in the order of microseconds. The cost of sending an additional byte of data ($\beta$), on the other hand, is about 0.1 nanosecond (indicating a bandwidth of about 12 GB/s).
The primary motivating factor for message aggregation is this discrepancy in the order of $\alpha$ and $\beta$ values. While $\alpha$ is in microseconds, $\beta$ is less than a nanosecond per byte, 
motivating the need for fine-grained messaging applications to aggregate many small messages into fewer large messages.

For clarity in usage of terms, we will use ``items" to refer to short messages the application wishes to send, and reserve ``message" to refer to aggregated data that the aggregation libraries send via the underlying communication mechanism. 
The use-cases for message aggregation range from all-to-all communication in MPI, where every rank wishes to send a relatively small number of items to every other rank, to streaming scenarios encountered in graph algorithms or in the well-known random-access benchmark from the HPC Challenge suite, where each worker continually generates a stream of items to others. While in the former case, one can anticipate an end point where the application signifies the contribution from a worker has ended, in the latter case, although such an end point may exist at some time, one must design the library for continuous flow of information. The distinction between the two cases is only quantitative, but may require different aggregation techniques. 
In contrast to both these, consider a parallel discrete event simulation with an optimistic (or speculative) protocol: here, items are events with a payload and a time-stamp associated with them. Destination (logical processes or agents) that receive items in time-stamp order are efficient, but out-of-order delivery leads to cascading rollbacks with high overhead. 

Observing the behavior of many such applications, and by virtue of the co-design of our new aggregation library with a graph algorithm, we inferred that, aside from effects on communication cost using message aggregation, a key metric that is influenced by message aggregation is latency. Different applications categories are differently affected by latency. We define latency here as the time from when an item is generated by the application to when it is delivered to the application on another processor. While aggregation reduces the cost of sending items, it increases the latency associated with each item. As items are buffered, latency is higher for messages than without any aggregation. In extreme cases, for applications that are highly latency sensitive, message aggregation may have to be avoided. But in practice, the large $\alpha$ cost of short messages means that we have to deal with the dual goals of reducing latency while simultaneously reducing the overhead.

The paper presents multiple schemes for aggregation in the context of SMP applications (where cores on a node are organized into multiple processes with multiple cores assigned to each), and compares them on the dual metrics of overhead and latency, for multiple applications represented by benchmarks. 
The schemes vary based on the level at which the aggregation is done: core/worker, or process, on both sending and receiving side. 
We analyze the benefits of SMP-aware message aggregation with different schemes, for different types of applications. The contributions of this paper are as follows.
\begin{itemize}
    \item We analyze message aggregation which is SMP-aware, with one buffer per destination node (or process), or one buffer per source node or both.
    \item We analyze the above schemes and how buffer sizes affect latency and overhead
    \item We describe a series of features that are engendered by application-co-design, which make the our library (called \tramlib{} ) versatile and responsive to the varying needs of the applications. 
\end{itemize}

We characterize latency and overhead of the different schemes considered using the following: We use two common short benchmarks, namely histogramming and index-gather, which allow isolated measurements of overhead and latency. We also demonstrate usefulness of our library on irregular applications including a graph applications and PDES, via benchmarks representative of those.

\section{Background}
The previous implementation of aggregation in Charm++ \cite{tramICPP14} focused on topology aware routing, aggregated items at PE or core level, and routed to the destination PE using a multi-dimensional routing scheme. In contrast, the new SMP-aware aggregation algorithm that we discuss in the paper, embodied in a library called \textit{tramlib},  aims to reduce latency of messages being aggregated. 

Our work is developed in the context of the Charm++ parallel programming system, and uses its features, including message-driven execution.
\charm{} divides work into units called chares. Overdecomposition, a key feature in ~\charm{}, allows multiple chares to be mapped to a processing element (PE), typically a core. Allowing multiple chares on a PE allows for computation and communication overlap. It also allows the ~\charm{} runtime to migrate chares to provide functionality like load balancing, shrink/expand, fault tolerance and others.

Shared memory parallelism with a large number of cores presents a new set of challenges for message aggregation. Existing systems like Active Pebbles do not support SMP to avoid dealing with overhead caused by atomics, locks and contention within node. 
Some systems like Conveyors~\cite{conveyors} make the assumption that communication within a process (with multithreading) will happen using shared memory and focus on messages sent from process-to-process. YGM~\cite{YGM} allows for coalescing of messages from source cores in a node or to destination cores in a node. It does so by performing node local sends and receives which rely on the efficient underlying implementation of local sends/receives instead of directly utilizing 
shared memory features.  The older TRAM system in Charm++ was also extended to support SMP  and multiple cores,  by assigning cores in a physical nodes as an extra dimension in its multi-dimensional routing scheme. However, grouping each core into a hyperplane does not help reduce internode messages. 
This was evidenced by the suboptimal performance in many SMP benchmarks and applications. Other works rely on posix shared memory~\cite{hybrid-mpi} for communicating between processes on the same physical node.

\section{Aggregation Schemes}
\label{sec:impl}

\subsection{SMP woes for fine-grained computations}
\label{sec:smp-issues}

In spite of the potential benefits of SMP mode, initial experiments in using message aggregation in SMP mode demonstrated a huge performance gap: SMP mode was more than 10 times slower than non-SMP in the simple histogramming benchmark of the Bale suite \cite{bale}. 

\begin{figure}[ht]
  \centering
  \includegraphics[width=0.5\linewidth]{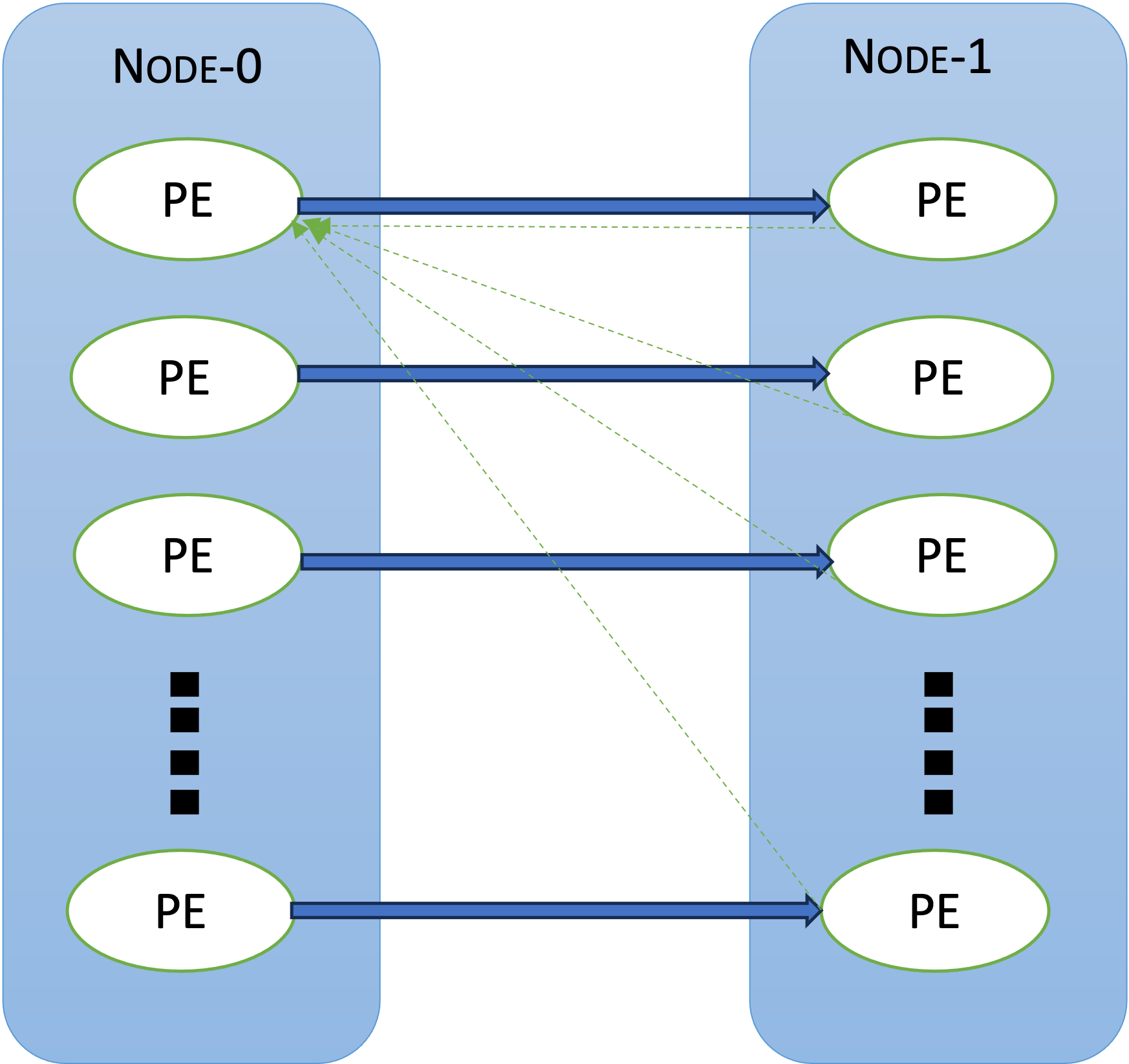}
  \caption{\pingack{} benchmark}
  \label{pingack}
\end{figure}

To analyze this issue, 
we wrote a simple benchmark called {\em \pingack{}}, which runs on two physical nodes. Each worker PE on the first physical node sends 1000 messages of a given size to the corresponding PE or core on the second physical node.  Each PE on the second node sends an ack to PE-0 after receiving {\em all} its messages. We measure the total time as time from the start of sends on the PE-0 to the time ack is received. Figure~\ref{pingack} shows the communication pattern between chares on PEs in the \pingack{} benchmark. With this benchmark, we expect that each PE on node-0 simultaneously sending 1000 messages to PEs on node-1 would create a heavy load for the underlying communication processing, but only for sending on node-0, and for receiving on node-1, allowing for better analysis. 

\begin{figure}[t!]
  \centering
  \includegraphics[width=0.75\linewidth]{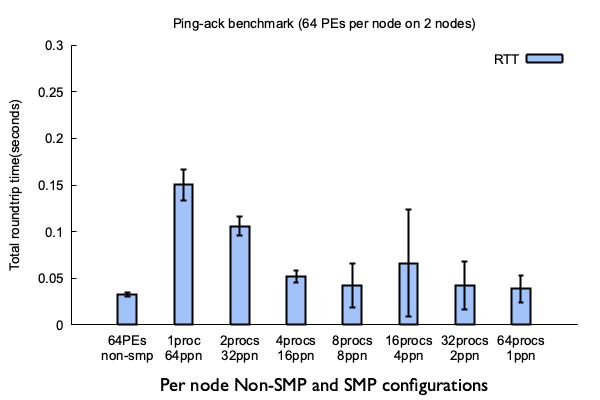}
  \caption{\pingack{} benchmark SMP (different process counts) vs non-SMP on 2 physical nodes}
  \label{multiprocess-1}
\end{figure}

Next, we compare the performance of \pingack{} benchmark in SMP and non-SMP mode in Figure~\ref{multiprocess-1},  performed on 2 physical nodes on the Delta supercomputer at NCSA. Multiple trials including initial warm-up ones are used to generate error bars. In the figure, the first and second histogram bars show the non-SMP version with 64 PEs and SMP with 64 PEs. The SMP version with 64 PEs is about $5\times$ slower that the non-SMP version! 

To understand this (and the remaining bars), note that Charm++ uses one dedicated core within each process for communication. This design has worked well in the broader context of ``normal" applications, which do significant computation for each word of communication. However, as confirmed with additional experiments using \pingack{}, for fine-grained applications, if the amount of work per word of communication was less than 167 nanoseconds, the communication thread (commThread) itself becomes a serializing bottleneck. As we use more processes per node, each with its own dedicated commThread, performance significantly improves (as seen in the figure). The number of messages sent by each PE on node 0 is adjusted with each configuration, so that the total number of messages from Node 0 to Node 1 remains the same. This alone was not sufficient: operating system daemons and activities such as gpgpu callbacks affect 1 core in each process, and in fine grained applications, 1 core needs to be set aside using explicit core-mapping primitives. With these two mechanisms, it becomes possible to overcome most of the communication related disadvantages of SMP mode for fine-grained applications. 

Recent work by Zambre et al.~\cite{zambre1,zambre2} has identified hindrances in the use of multiple NICs and concurrency in communication as an additional issue. Using many processes per node also partially mitigates those concerns. 
Although we have identified additional optimizations for enhancing communication performance of SMP mode, the above suffice for our current purpose.  

\subsection{Basic Aggregation Schemes for aggregation in SMP context}
\begin{figure}[ht]
  \centering
  \includegraphics[width=0.75\linewidth]{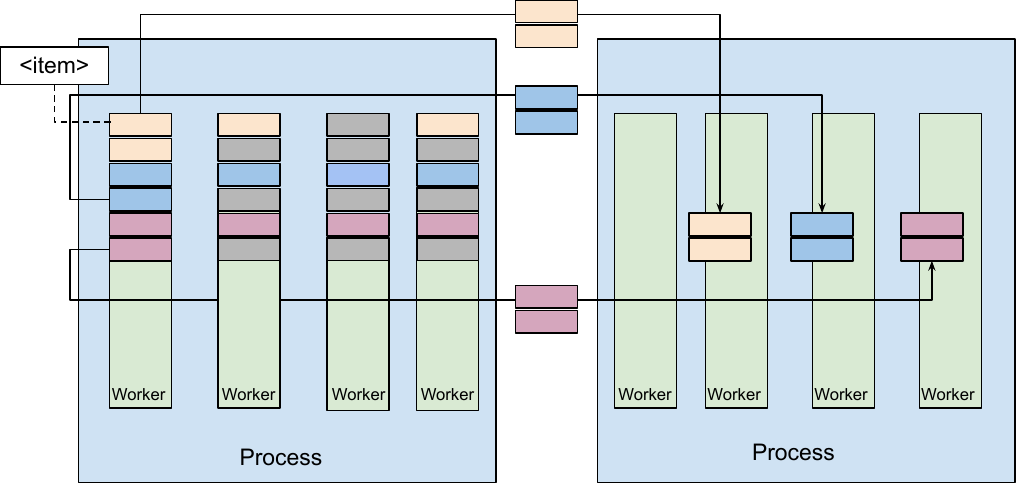}
  \caption{\pp{}: Source worker maintains a buffer per destination worker}
  \label{fig:ww}
\end{figure}

The basic scheme is one in which the aggregation is done at the level of a worker (i.e. a PE in charm terminology, corresponding to a Pthread bound to a core)  at both source and destination, shown in Figure~\ref{fig:ww}. Concretely, if there are w workers, each worker maintains w-1 buffers for items going to every other worker. A buffer is sent to the destination as a message when it is full or when the application asks \libtram{} to flush accumulated items. This scheme, denoted \pp{}, does not leverage the shared memory across workers. 

\begin{figure}[ht]
  \centering
  \includegraphics[width=0.75\linewidth]{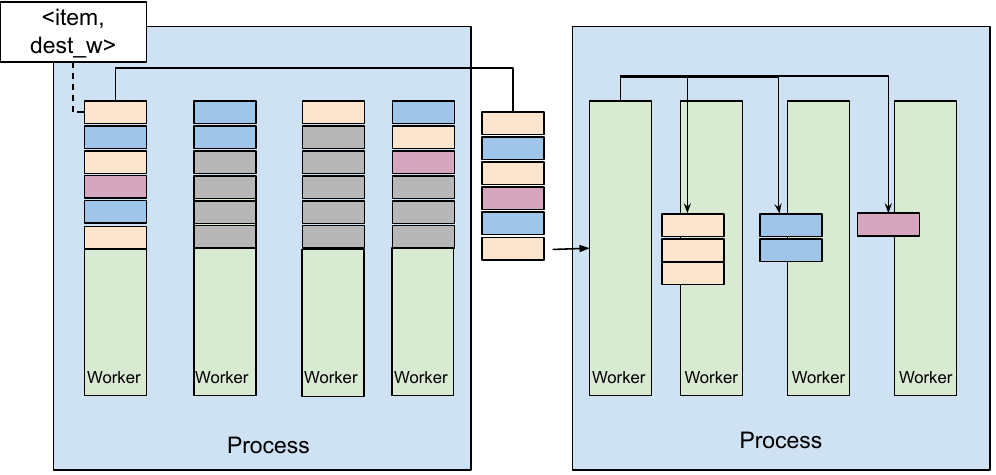}
  \caption{\pns{}: Source worker maintains a buffer per destination $process$}
  \label{fig:wps}
\end{figure}

\begin{figure}[ht]
  \centering
  \includegraphics[width=0.75\linewidth]{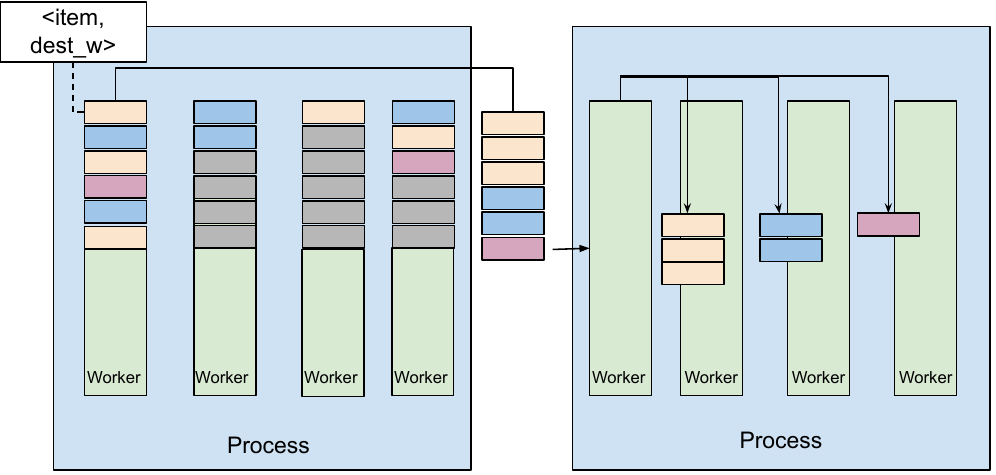}
  \caption{\psn{}: Source worker maintains a buffer per destination $process$}
  \label{fig:wsp}
\end{figure}

In the second scheme, each worker aggregates messages  by destination process (maintaining a buffer for each of the p processes), shown in Figure~\ref{fig:wps}. Since the items need to be delivered to individual workers, it is necessary to sort (or group) items within the message at some stage. We  considered two variants for this purpose. In the first variant, the sorting (denoted by $s$) or grouping of items is performed at the destination process (Figure~\ref{fig:wps}), denoted by \pns{}. In the other variant, shown in Figure~\ref{fig:wsp}, the sorting or grouping is performed at the source worker, denoted by \psn{}. 

\begin{figure}[ht]
  \centering
  \includegraphics[width=0.75\linewidth]{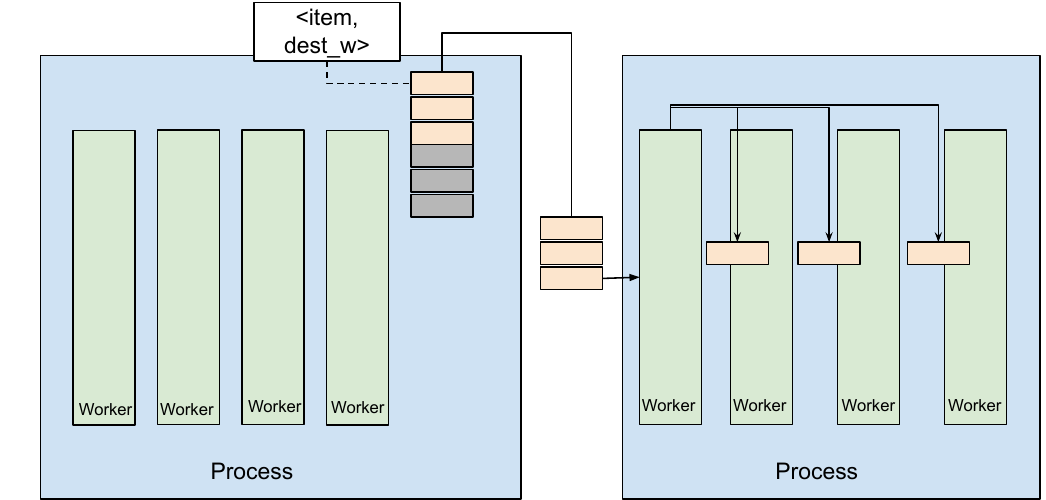}
  \caption{\nn{}: Source $process$ maintains a buffer per destination $process$}
  \label{fig:pp}
\end{figure}

In the next scheme, denoted \nn{},  we aggregate messages on each source process, by destination process. I.e. on each {\em process} there is only one buffer for each target process to which all the workers within the process contribute. This coalescing in the source process is achieved using atomics. This is illustrated in Figure~\ref{fig:pp}.

The same grouping techniques can be extended one level up to the physical node, if it houses multiple processes. 
Also, in the past, multi-dimensional routing across nodes had been considered, where messages going from one physical node to another, go via one or more intermediary nodes, which act like hubs in a routing network and help reduce overhead further. 
Although practically useful, we omit the discussion of these higher level aggregation schemes from this paper, as they are orthogonal to the SMP considerations that are the focus of this paper.

In all of the schemes, a few basic optimizations are applied: messages sent by flush operations are resized, so that empty portions of the aggregation buffer are not sent. Buffers can be flushed, optionally, when the processor is idle, or when triggered by the application, or by a timeout. Charm++'s expedited methods are used to prioritize \tramlib{} messages over other ordinary application messages.

The aggregation capability is implemented as a ~\charm{} library. At initialization time, the user code passes a pointer to the charm++ object and function to which data needs to be delivered.

When items are inserted, \tramlib{} checks if data item count for the destination has reached the buffer size. If so, the buffer of aggregated data items is sent to the receiver chare in \tramlib{}. The receiver then returns the data to the application. There are few variations in this implementation depending on the type of scheme.

\subsection{Analysis}
Below we analyze the following costs associated with node-aware \tramlib{}:
\begin{enumerate}
    \item Memory overhead
    \item Number of messages sent
    \item Message send cost
    \item Processing delays at receiver
\end{enumerate}
\textbf{Memory overhead:} We use $g$ to denote number of items in TRAM buffer and $m$ to denote the size of each item, similar to notation used in previous work TRAM~\cite{tramICPP14}. With node-aware \tramlib{}, we do not use topological multi-dimensional routing of previous TRAM. So, given the total number of processes $N$, and assuming $t$ workers per process (i.e. Pthreads bound to disjoint cores), the memory overheads related to buffer allocation for the different schemes are:
\begin{itemize}
    \item \pp{}: Aggregating messages on each source PE per destination PE = $g*m*N*t$ per core or $g*m*N*t^2$ per process
    \item \pns{}, \psn{}: Aggregating messages on each source PE per destination node = $g*m*N$ per core or $g*m*N*t$ per process

    \item \nn{}:  Aggregating messages at each source process (instead of PE) per destination process = $g*m*N$ per source process
\end{itemize}

However, the overall impact of these alternatives is complicated by the fact that the optimal buffer size $g$ would be different for each of the schemes, and also because for some applications, latency of individual messages may be important. 

\textbf{Number of messages sent:} For a buffer size $g$, if $z$ items are send from each source PE, the number of messages sent for different schemes is:
\begin{itemize}
    \item \pp{}: The number of messages sent per source PE will have a lower bound of $z/g$ and an upper bound of $z/g + N*t$
    \item \pns{}, \psn{}: The number of messages sent per source PE will have a lower bound of $z/g$ and an upper bound of $z/g + N$
    \item \nn{}: With source process aggregation, messages sent per source process will have a lower bound of $z/g$ and an upper bound of $z/g + N$
\end{itemize}
This assumes no intermittent flushing is triggered by the application except once at the end. The lower bound is achieved when partially filled buffers have to be flushed at the end of the communication phase. It can be seen that for long streaming applications, where $z>>g$, the second term in the upper bound is vanishingly small, and so the schemes are almost identical in number of messages sent. But for short streams of items, such as in a short all-to-all operation, aggregating by the destination process (i.e. using one buffer for all items going to different destination PEs belonging to the same destination process) has the least number of messages. 

\textbf{Message send cost:}
Assuming alpha-beta model, sending z items would take $z*(\alpha + \beta*b)$ for communication cost for sending each item as a message, assuming $b$ bytes per item are sent. With coalescing, provided buffer size is $g$, the communication cost for $z$ items is $(z/g)*(\alpha + \beta*b*g)$ or $(z/g)*\alpha + \beta*b*z$, essentially reducing the $\alpha$ component in the message cost by $g$. This has implications on latency however, where now the average message latency is increased based on buffer fill rate. Assuming buffer fill rate is $r$, the latency of an item in the buffer can increase by up to $g/r$. However, on the other hand, in some cases with several tiny messages, aggregating messages can also improve latency of messages since the sender PE is blocked less for future messages, hence improving overall latencies of messages. Smaller buffer sizes and node-aware schemes can further improve latencies, for latency-sensitive applications.

\textbf{Processing delays:} When coalescing messages, there are a few delays like higher cost per byte and buffer fill delays. Another delay is the overhead $O$ that is added once per aggregated message sent or received. On the sending processor, the overhead is from contention when we maintain common buffers per physical node. At destination node or process, the receiving PE sorts or groups messages by PE and communicates the coalesced message locally to each PE. The sorting or grouping of items by worker, for $t$ workers per process, results in an additional delay of $O(g+t)$ to output a grouped coalesced message of the same size as the buffer $g$. Local sends are typically fast in shared memory systems like ~\charm{}.

\subsection{Benchmarks and Applications}
All applications are written in ~\charm{} and use the API from the~\tramlib{} aggregation library implemented as a ~\charm{} library module.

\textbf{Histogram}: The histogram benchmark is based on the one from the Bale benchmark suite \cite{bale}. It uses a distributed histogram table on all PEs, and all PEs send a given number of updates to the distributed histogram. Each PE invokes the flush call at the end of all updates to flush any remaining updates in the ~\tramlib{} buffers. Since there are no dependent communication items created as a result of an histogram update sent, latency is not directly relevant this benchmark. Instead, it can be used to measure overhead, in isolation. 

\textbf{Index-gather (IG)}: This example is also based on Bale benchmark suite. Each PE sends a given number of requests to other PEs which send a response upon receiving the request. This benchmark helps us measure latency for a message since we can measure the time between a request sent and response received on a given PE. While latency can be measured in other benchmarks as well, measuring time between sends and receives on different nodes can include clock skew error. Hence we use IG for analyzing latency for various aggregation schemes.

\textbf{SSSP}: The single source shortest path graph algorithm used in this work has vertices distributed across chares, with one chare per PE. The algorithm performs speculative execution, by computing distances based on available distances of neighbors of vertices to the source vertex. If a PE receives an updated distance for a vertex that is smaller than previously known distance, the value is updated. The updated value is sent to neighbor vertices of this vertex, which then may update their distance similarly. The neighbor vertices maybe on remote PEs or local PEs. A threshold value is used in this benchmark that helps prioritize updates with smaller distances in order to minimize wasted updates. This is because updates with larger updates are likely to become wasted updates since they are likely to updated with smaller values later. The program terminates when all updates have been consumed. This benchmark is latency sensitive since higher latency of messages is likely to result in more wasted updates being created.

\textbf{PDES - optimistic}: This refers to the optimistic concurrency control engine in parallel discrete event simulation (PDES). Similar to SSSP, this benchmark will help us analyze rollbacks resulting from higher latency in PP compared to the node-aware schemes. PHOLD is a well-known synthetic benchmarks used in PDES. We implemented a synthetic PHOLD with a place-holder simulation engine to analyze how latency affects rollbacks. In this engine, we do not perform real rollbacks; instead we only keep track of out-of-order messages received.
\label{sec:eval}
\section{Evaluation}

\subsection{System and benchmarks}
Our experiments are performed on the Delta supercomputer at NCSA. We use the SMP build of ~\charm{}  for our runs with OFI as the network layer. Each Delta node is an AMD EPYC 776, dual-socket with 128 cores per node, 1 hardware thread per core with a clock rate of 2.45GHz and 256 GB RAM. 
We use 8 processes per physical node, to allow for power of 2 processes, since we need 1 core for communication thread per process. Remaining cores are left unused. Hence on each Delta node, we have 64 worker cores distributed across 8 processes (with 8 ppn each) and 8 communication thread cores.

\begin{figure}[ht]
  \centering
  \includegraphics[width=0.85\linewidth]{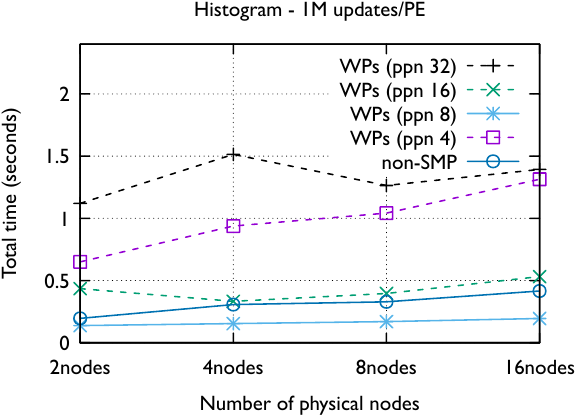}
  \caption{Histogram 1M smp vs non-smp}
  \label{fig:comm-histo}
\end{figure}

Our initial experiment is to address the communication thread bottleneck in SMP mode, detailed in section ~\ref{sec:smp-issues}. For our purposes, we use multiple processes per node to mitigate communication thread performance issues. Varying the number of worker threads per process for histogram benchmark, in Figure~\ref{fig:comm-histo}, we observe that setting 8 workers or ppn for smp process in WPs scheme performs on par with the non-smp implementation for histogram benchmark. For the remaining experiments, we use ppn of 8 on Delta, with 8 process per node.

\begin{figure}[ht]
  \centering
  \includegraphics[width=0.85\linewidth]{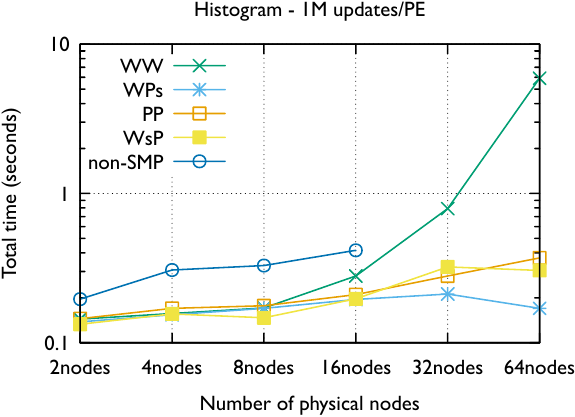}
  \caption{Histogram 1M example}
  \label{fig:histo-eg1}
\end{figure}

To assess the effectiveness of our smp-aware aggregation schemes, in Figure~\ref{fig:histo-eg1}, we measure overall execution time for different schemes with increasing node counts for histogram with 1 million updates per PE (or worker). This is an example of weak scaling as the amount of work per PE (1 million updates) remains the same. We observe that the \pns{} scheme scales well for all 64 nodes.~\nn{} and \psn{} also scale till 64 nodes, but \nn{} has overheads from atomics and in case of \psn{} which performs sorting or grouping before sends, we observe poorer scaling than \pns{}.~\pp{} on the other hand stops scaling after 16 nodes. This is likely due to the buffer size of 1024, being not filled up per worker at 32 nodes onwards. For 32 nodes, with 8 processes each with 8 worker, the total number of destinations per source worker is 2048, which means for an update count of 1 million, each destination worker would receive only 500 values. Therefore, the sends are dominated by flush costs resulting in 2048 messages per source worker compared to a flush for \nn{}, say, requiring only 32 messages.

\begin{figure}[ht!]
  \centering
  \includegraphics[width=0.85\linewidth]{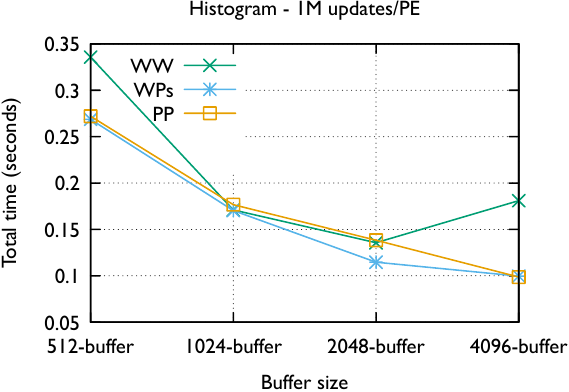}
  \caption{Histogram 1M example - varying buffer size for 8 node runs}
  \label{fig:histo-buf}
\end{figure}

We briefly analyze effects of buffer size, in Figure~\ref{fig:histo-buf} and find that the node-level aggregation schemes perform better with increasing buffer sizes for histogramming. \pp{} requires flushes at 4k buffer size for 1M updates per PE, hence overall time gets worse with increase in buffer size beyond 2k.

\begin{figure}[ht]
  \centering
  \includegraphics[width=0.85\linewidth]{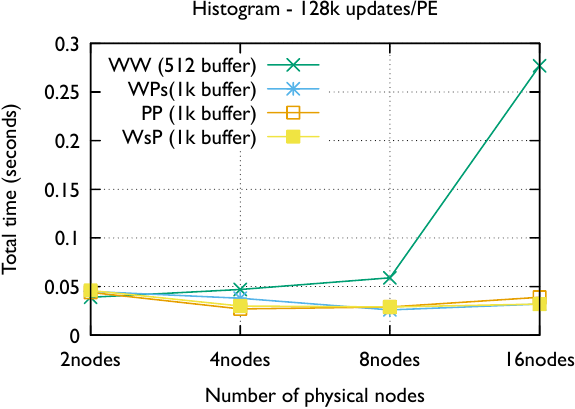}
  \caption{Histo 128K updates/PE, higher costs from flushes for \pp{} for 8 nodes and above}
  \label{fig:histo128k-ff}
\end{figure}

We can think of applications, on one end being streaming applications and on the other end latency sensitive applications that have fewer updates requiring frequent flushes. To mimic applications with frequent flushes, we analyze performance of histograming benchmark with only 128K updates generated per PE. This is shown in Figure~\ref{fig:histo128k-ff}.

For very small updates, eg. 128k updates/PE, the destination PE buffers do not fill up in case of PP for 8 nodes onwards, hence the execution time is dominated by flush calls. Since flush would empty all destination buffers, the total number of messages in PP is higher than for \pns{} and \nn{} where the number of buffers is reduced by a factor of PEs per process. In figure~\ref{fig:histo128k-ff}, the difference in overall time is significant between schemes with \pns{} and \nn{} performing significantly better. \nn{} likely incurs overhead of atomics, hence overall execution time does not improve compared to \pns{}. We have tuned the buffer size depending on application performance and latency in this set of experiments, with a buffer size of 512 for WW and 1024 for the other schemes.

\begin{figure}[h]
  \centering
  \includegraphics[width=0.85\linewidth]{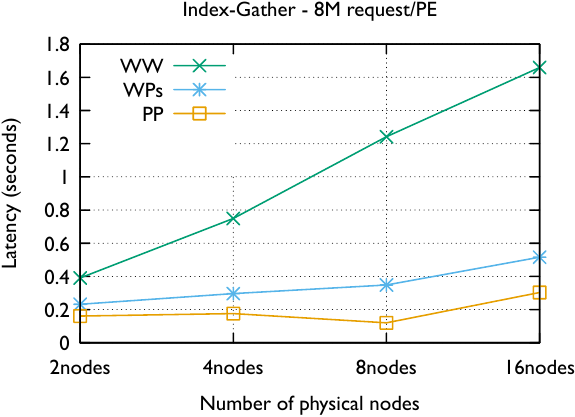}
  \caption{IG 8m - latency}
  \label{fig:ig-lat}
\end{figure}

\begin{figure}[!h]
  \centering
  \includegraphics[width=0.85\linewidth]{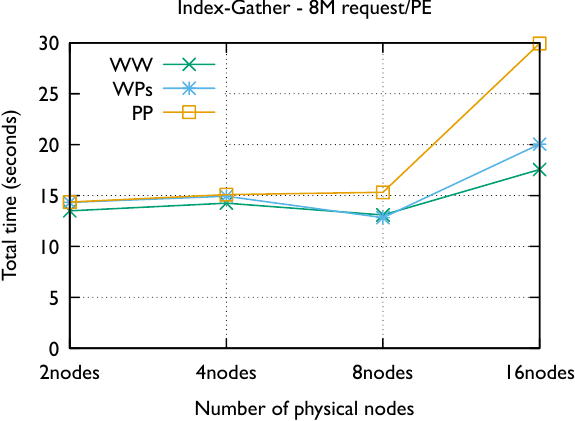}
  \caption{IG 8M - total time}
  \label{fig:ig-overall}
\end{figure}

As mentioned before, latency is a key metric in many applications. It can be affected by buffer size, overhead of message cost or aggregation scheme used. We measure latency using the IG benchmark, using a buffer size of 1024 for all the aggregation schemes. Figure~\ref{fig:ig-lat} shows that latency of \nn{} \textless ~\pns{} \textless ~\pp{}. For overall cost, however, for IG, in Figure~\ref{fig:ig-overall}, the overhead of sorting in \pns{} and the overhead of atomics in \nn{}'s seems to affect overall execution time at 16 nodes. However, for IG, we are mainly concerned with using the benchmark to measure latency of different schemes, which we will later apply to other applications.


\begin{figure}[ht]
  \centering
  \includegraphics[width=0.85\linewidth]{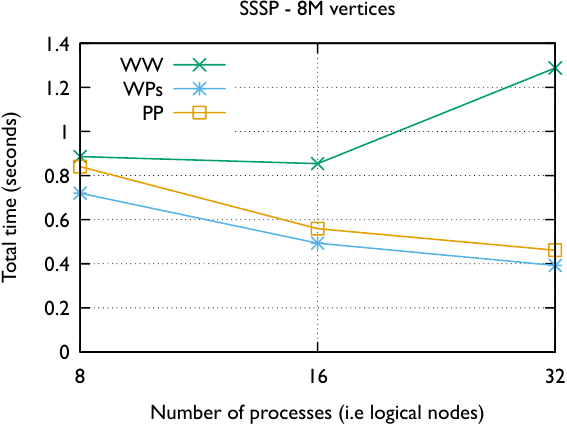}
  \caption{SSSP - small problem size - Time}
  \label{fig:sss-small-input}
\end{figure}

\begin{figure}[ht]
  \centering
  \includegraphics[width=0.85\linewidth]{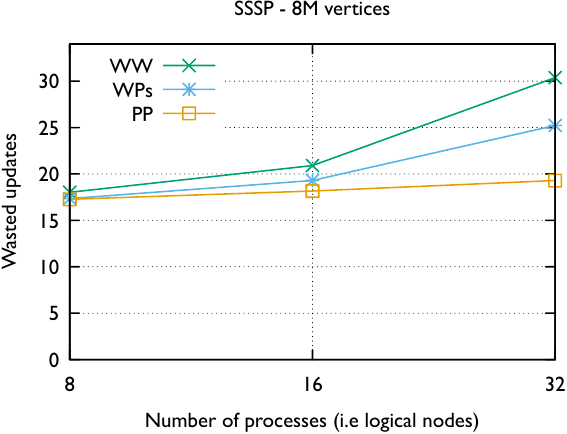}
  \caption{SSSP - small problem size - Wasted updates(normalized)}
  \label{fig:sss-small-input-wu}
\end{figure}

For latency-sensitive applications such as SSSP, with smaller problem sizes, more PEs also mean higher chances of wasted updates while PEs wait on updates from other PEs. We show how different schemes affect wasted updates for small problem size in Figures~\ref{fig:sss-small-input} and~\ref{fig:sss-small-input-wu}. In terms of wasted updates, we observe that ~\nn{} \textless ~\pns{} \textless ~\pp{}.

\begin{figure}[ht]
  \centering
  \includegraphics[width=0.85\linewidth]{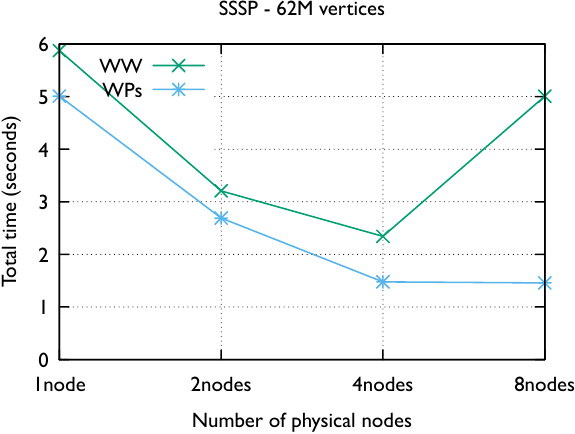}
  \caption{SSSP - large input size example, time}
  \label{fig:sssp-large-input}
\end{figure}

\begin{figure}[ht]
  \centering
  \includegraphics[width=0.85\linewidth]{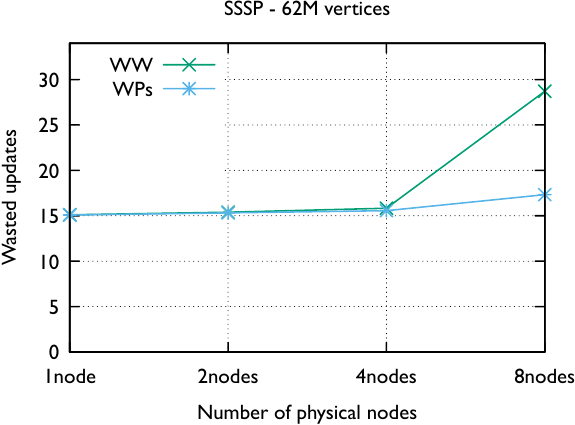}
  \caption{SSSP - large input size example, wasted updates (normalized)}
  \label{fig:sssp-large-input-wu}
\end{figure}

For SSSP with large problem sizes, which scale well across multiple nodes, we do not see a significant difference in wasted updates across different schemes as seen in Figure~\ref{fig:sssp-large-input} amd ~\ref{fig:sssp-large-input-wu}. However, \pns{} performs considerably better than \pp{}, which is likely from frequent flush calls and memory footprint. We plan to explore this further using PAPI performance counters and other analysis techniques.

\begin{figure}[ht]
  \centering
  \includegraphics[width=0.85\linewidth]{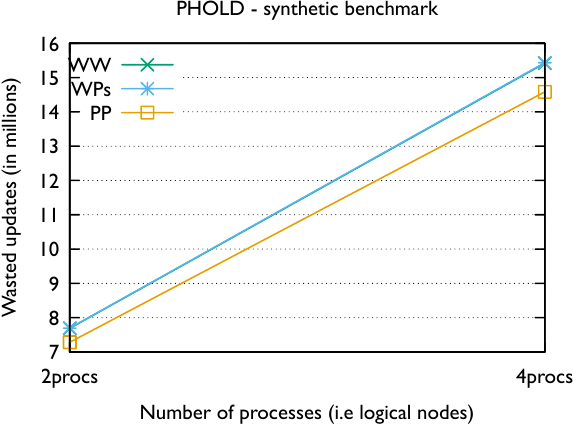}
  \caption{PHOLD synthetic - Wasted updates(in millions)}
  \label{fig:phold-wu}
\end{figure}

Finally, for PHOLD synthetic benchmark, we run our experiments with a higher ppn of 32 to see benefits of message consolidation in node-aware schemes compared to node-unaware aggregation schemes. The PHOLD benchmark's rejected updates are more than 5\% fewer for node-aware~\nn{} scheme, in Figure~\ref{fig:phold-wu}. For PHOLD,~\pp{}'s execution time was much higher (over 5x) compared to other schemes. This could be due to inefficient and frequent flush operations called from the synthetic benchmark, which we are further investigating. The greater than 5\% improved wasted updates for ~\nn{} can be significant for PDES simulations where rollbacks are expensive.

\section{Related Work}
\label{sec:related}
Several works have shown the benefits of message aggregation.
Early work was in the context of implementation of all-to-all collective by Thakur and Choudhary\cite{alltoall}. Sameer Kumar et. al~\cite{CommLib02} studied performance of all-to-all messaging with routing and aggregation in 2-D mesh, 3-D grid and hypercube virtual topologies. In contrast to all-to-all, where all data is handed over to the library at once at the beginning, streaming applications need to support continuous stream of short messages. 
TRAM~\cite{tramICPP14} in ~\charm{} supports streaming as well as all-to-all, performing routing through an N-dimensional virtual topology
and dynamically aggregating messages to reduce message count. The all-to-all performance from TRAM paper shows up to an order improvement in performance from routing and aggregation. Additionally, in a run on 64 nodes, increasing the buffer size up to a certain size, here 8K bytes, is best performing, while the best dimensionality for virtual topology can be seen to depend on data size contributed for each destination. The minimum latency of an item remains the same, but the number of
messages is reduced by the size of the buffer.
Active Pebbles~\cite{active-pebbles} is a parallel programming model for data-
driven applications. It aggregates messages meant for a destination process (on a destination node), performs local reductions if 
needed and routes the messages through a virtual topology. Active Pebbles however does not support a shared memory programming 
model to avoid within node contention.

Shared memory parallelism with large number of cores presents a new set of challenges for message aggregation. Existing system like Active Pebbles do not support SMP to avoid dealing with atomics, locks and contention within node. Some 
systems like Conveyors~\cite{conveyors} make the assumption that communication within a process (with multithreading) will happen using shared memory and focus on messages sent from process-to-process. YGM~\cite{YGM} allows for coalescing of messages from source processes, mapped to cores in MPI-everywhere model, in a node or to destination processes, mapped to cores, in a destination node. It does so by performing node local sends and receives which relies on the efficient underlying implementation of local sends/receives instead of directly utilizing SMP constructs.  In TRAM, to support SMP cores, assigning physical nodes as a dimension would still require routing/aggregation within a node and add an extra dimension, leading to sub-optimal performance. Other works rely on posix shared memory~\cite{hybrid-mpi} for communicating between processes on the same physical node.
Given the SMP advantage outlined earlier (section~\ref{sec:intro}), we aim to overcome these challenges and support node-aware message aggregation within SMP context. 

\section{Conclusion and Future Work}
\label{sec:concl}
We have discussed the design of various message aggregation schemes suited for today's large supercomputer nodes. We have demonstrated the effectiveness of SMP-aware aggregation schemes that utilize process-local sends and atomics to coalesce messages in a shared memory setting. We analyzed the schemes using two key metrics, namely overhead and latency. While message aggregation aims to lower overall execution time by reducing communication cost, aggregation increases latency of items being aggregated. For latency sensitive applications, we have shown  benefits resulting from lower latency aggregation schemes. In the future, we plan to support prioritization of items, which should help latency or cost sensitive applications such SSSP and PDES even more directly. 
Also, we plan on further analyzing benefits of node-aware aggregation across processes within a physical node, for instance using XPMEM for across process shared memory benefits, in addition to within-process message aggregation.

\bibliographystyle{IEEEtran}
\bibliography{citations,group}
\vspace{12pt}

\end{document}